\documentclass[usenatbib]{mn2e}
\usepackage{amsmath}
\usepackage{natbib}
\usepackage{epsfig}
\usepackage{txfonts}

\newcommand{\nbb}{{n^{\rm pl}}}

\newcommand{\vgh}{{\hat{\boldsymbol\gamma}}}
\newcommand{\vghp}{{\hat{\boldsymbol\gamma}'}}

\newcommand{\vbh}{{\boldsymbol{\hat{\beta}}}}

\newcommand{\xc}{x_{\rm c}}

\newcommand{\id}{{\,\rm d}}

\newcommand{\beq}{\begin{equation}}   %

\newcommand{\eeq}{\end{equation}}   %

\newcommand{\beqa}{\begin{eqnarray}}   %

\newcommand{\eeqa}{\end{eqnarray}}   %

\newcommand{\beal}{\begin{align}}

\newcommand{\enal}{\end{align}}

\newcommand{\bspl}{\begin{split}}

\newcommand{\espl}{\end{split}}

\newcommand{\bsub}{\begin{subequations}}

\newcommand{\esub}{\end{subequations}}

\newcommand{\bmulti}{\begin{multline}}   %

\newcommand{\beqm}{\begin{mathletters}}   %

\newcommand{\eeqm}{\end{mathletters}}   %

\newcommand{\me}{m_{\rm e}}

\newcommand{\Ne}{N_{\rm e}}

\newcommand{\Te}{T_{\rm e}}

\newcommand{\Tg}{T_{\gamma}}

\newcommand{\The}{\theta_{\rm e}}

\newcommand{\sigT}{\sigma_{\rm T}}

\newcommand{\vek} [1]{\mbox{\boldmath${#1}$\unboldmath}}

\newcommand{\pot}[2]{#1 \times 10^{#2}}

\newcommand{\musc}{\mu_{\rm sc}}

\newcommand{\GHz}{{\rm GHz}}

\newcommand{\expf}[1]{{{\rm e}^{#1}}}

\newcommand{\MJy}{{\rm MJy}}
\newcommand{\sr}{{\rm sr}}
\newcommand{\keV}{{\rm keV}}

\usepackage{color}

\renewcommand{\nbb}{{n^{\rm Pl}}}

\usepackage{hyperref}
\usepackage{grffile}
\usepackage{graphics}
\usepackage{booktabs}

\title[Multiple scattering SZ]
{Multiple scattering Sunyaev-Zeldovich signal I: lowest order effect}

\author[Chluba, Dai, \& Kamionkowski]{
J.~Chluba\thanks{E-mail: jchluba@pha.jhu.edu},
L. Dai\thanks{E-mail:ldai@pha.jhu.edu} and M.~Kamionkowski\thanks{E-mail: kamion@jhu.edu} 
\\
Department of Physics and Astronomy, 
Johns Hopkins University, Bloomberg Center, 
3400 N. Charles St., Baltimore, MD 21218, USA
}

\begin{document}

\date{Accepted 2013 September 30; Received 2013 August 26}

\maketitle

\begin{abstract}
Future high-resolution, high-sensitivity Sunyaev-Zeldovich (SZ) observations of individual clusters will allow us to study the dynamical state of the intra cluster medium (ICM). To exploit the full potential of this new observational window, it is crucial to understand the origin of different contributions to the total SZ signal. Here, we investigate the signature caused by multiple scatterings at lowest order of the electron temperature. Previous analytic discussions of this problem used the isotropic scattering approximation (ISA), which even for the simplest cluster geometries is rather rough. We take a step forward and consistently treat the anisotropy of the ambient radiation field caused by the first scattering. 
We show that the multiple scattering SZ signal directly {\it probes} line of sight {\it anisotropies} of the ICM, thereby delivering a new set of SZ observables which could be used for 3D cluster-profile reconstruction. The multiple scattering signal should furthermore correlate spatially with the cluster's X-ray and SZ polarization signals, an effect that could allow enhancing the detectability of this contribution.
\end{abstract}


\section{Introduction}
\label{sec:Intro}
A number of high-resolution Sunyaev-Zeldovich (SZ) experiments,
including ALMA\footnote{Atacama Large Millimeter/submillimeter Array},
CARMA\footnote{Combined Array for Research in Millimeter-wave
Astronomy}, CCAT\footnote{Cornell Caltech Atacama Telescope},
and MUSTANG\footnote{MUltiplexed Squid TES Array at Ninety GHz},
are underway or planned, promising a dramatic increase in
sensitivities, spatial resolution and spectral coverage over
the next few years. These experiments will allow us to directly
probe the dynamical state of the intra cluster medium (ICM) of individual clusters,
opening a new window to understanding their cosmological
formation and evolution. To realize the full potential of these
observations, it is important to understand the SZ signal and its
dependence on the cluster atmosphere with high precision.

Since the first considerations of the thermal SZ (thSZ) effect \citep{Zeldovich1969} and kinetic SZ (kSZ) effect \citep{Sunyaev1980}, several related signatures have been discussed in the literature. For instance, because the electrons residing inside clusters are very hot [$k\Te\simeq 5\,\keV-10\,\keV$ (with $k$ denoting the Boltzmann constant and $\Te$ the electron temperature)] relativistic temperature corrections need to be included \citep{Wright1979, Rephaeli1995, Challinor1998, Itoh98, Sazonov1998}. Also, higher order kSZ terms, both caused by the cluster's bulk motion \citep{Sazonov1998, Nozawa1998SZ, Challinor1999} and the motion of the observer \citep{Chluba2005b, Nozawa2005} arise\footnote{The interpretation and origin of higher order kSZ terms was recently clarified by \citet[][CNSN hereafter]{ChlubaSZpack}. Both relativistic temperature and kinematic SZ correction can be efficiently modeled with high precision over a wide range of parameters using {\sc SZpack} (see CNSN). }. Furthermore, relativistic electrons can create a non-thermal SZ signature indicative of past AGN activity and merger shocks \citep{Ensslin2000, Colafrancesco2003}.
Additional effects are related to details of the cluster's atmosphere. For example, internal bulk and turbulent gas motions introduce spatially varying SZ signatures \citep[e.g.,][]{Chluba2002, Nagai2003, Diego2003}. Line-of-sight variations of the electron temperature cause a non-trivial SZ morphology which depends on  {\it moments} of the cluster temperature and velocity profiles \citep[][CSNN hereafter]{Chluba2012moments}. Extra pressure support from large-scale magnetic field deforms the electron distribution inside clusters \citep{Koch2003}.  
And finally, shocks and substructures in merging clusters leave distinct signatures in the SZ flux \citep[e.g.,][]{Komatsu2001, Kitayama2004, Colafrancesco2011, Korngut2011, ElGordo2012, Mroczkowski2012, Zemcov2012, Prokhorov2012}. 

In this work we consider the effect of multiple scattering on the SZ signal \citep{Sunyaev1980, Sazonov1999, Molnar1999, Dolgov2001, Itoh2001, Colafrancesco2003, Shimon2004}. As we show here, this signature depends on inhomogeneities in the cluster's electron and temperature distribution. It therefore provides a new set of SZ observables that are valuable for the reconstruction of 3D cluster profiles\footnote{See \citet{Morandi2012} for recent demonstration based on future high-precision measurements of the thSZ effect using CCAT.}, potentially allowing geometric degeneracies to be broken. We restrict our analysis to the lowest order in the electron temperature. Higher order corrections will be discussed in \citet[][CD13 hereafter]{Chluba2013prep}.

The main new physical ingredient is that for the second scattering we include the full {\it anisotropy} of the singly-scattered radiation field. In previous studies \citep[e.g.,][]{Itoh2001, Colafrancesco2003, Shimon2004}, this aspect was omitted; however, even for the simplest cluster geometries this approximation is very rough (see Sect.~\ref{sec:multiple_scatt_lowest}).
Some discussion of the scattering-induced anisotropy in the radiation field can be found in connection with SZ polarization effects \citep[e.g.,][]{Sunyaev1980, Sazonov1999, Lavaux2004, Shimon2006} and accretion flows \citep[e.g.,][]{Sunyaev1985}, but the SZ signature in the frequency dependence of the specific intensity that we consider here was previously not studied.
We determine a new set of SZ observables related to the second scattering correction (Sect.~\ref{sec:total_signal_def}) and demonstrate how the signal depends on the cluster's geometry (see Fig.~\ref{fig:total_corr_lowest}).
We also find that the spatial pattern of the second scattering SZ signal can be used to enhance its detectability by cross-correlating with X-ray and future SZ polarization data (Sect.~\ref{sec:total_corr_spatial}).

For those readers less interested in the details of the derivation, we note that
the central result of the paper is an expression,
Eq.~\eqref{eq:total_SZ_signal_r}, for the total change
$\Delta I(x,\vgh)$ to the specific intensity induced
along a line of sight $\vgh$ through the cluster in the first and
second scattering as a function of frequency $x$.
The new parameters $\gamma_{\rm T}(\vgh)$ and $\gamma_{\rm E}(\vgh)$ that appear
in that expression are explicitly given for a spherically symmetric
cluster as a function of impact parameter $b$ in
Eq.~\eqref{eq:gammaEspherical}.

\section{Equation of Radiative Transfer}
\label{sec:radiativetransfer}
We begin with the radiative transfer equation in an anisotropic, static medium\footnote{The left-hand side (the Liouville operator) of the Boltzmann equation reads $c^{-1}{\rm d} n_\nu/{\rm d} t=c^{-1}\partial_t n_\nu+\vgh \cdot \nabla n_\nu$ ($c$ being the speed of light), thus `static' mean that changes in the medium are happening at time-scales much longer than the photon travel time through the ICM. In this case, the solution for the photon field becomes stationary (no explicit dependence on time) and the time-derivative term can be neglected, such that $c^{-1}{\rm d} n_\nu/{\rm d} t\approx \vgh \cdot \nabla n_\nu = \partial_s n_\nu$, where $s$ parametrizes the distance along the photon path \citep[cf. \S6.4 of][]{Mihalas1984}. In the cluster frame, this condition is fulfilled even if the cluster is moving at constant velocity.} \citep[e.g.,][]{Mihalas1978, Rybicki1979, Mihalas1984}:
\beal
\label{eq:Boltzmann}
\frac{\partial n(x, \vek{r}, \vgh)}{\partial \tau}= \mathcal{C}[n(x, \vek{r}, \vgh)].
\end{align}
Here, $n(x, \vek{r}, \vgh)$ denotes the photon occupation number at
the location $\vek{r}$, in the direction $\vgh$ and at frequency
$x=h\nu(z)/k \Tg(z)={\rm const}$, where $\Tg=T_0(1+z)$ is the cosmic microwave background (CMB) temperature at redshift $z$ with the present-day monopole $T_0=2.726\,\rm K$ \citep{Fixsen1996, Fixsen2002, Fixsen2009}, and $h$ denotes the Planck constant.
The radiative transfer equation is solved along the photon
path (parallel to $\vgh$) through the ICM, parametrized by the Thomson optical depth,
$\tau(\vgh)=\int \sigT \Ne(\vek{r}) \id l$, where $\Ne(\vek{r})$ is the free electron number density.

\subsection{Compton collision term at the lowest order in $k\Te/\me c^2$}
The collision term $\mathcal{C}[n]$ in Eq.~\eqref{eq:Boltzmann} accounts for Compton scattering of photons of frequency $x$ and direction $\vgh$ out of the beam and for scattering of photons of other directions and frequencies into the beam.  To linear order in the electron-temperature parameter $\The \equiv k \Te/\me c^2$ (where $\me$ is the electron mass), the collision operator for anisotropic incoming radiation, as shown in
Appendix~\ref{app:coll_term}, is \citep{Chluba2012}
\beal
\label{eq:gen_coll_Te}
\mathcal{C}[n] &\approx \frac{3}{16\pi} \int \id^2\vghp (1+\musc^2) [n(x, \vek{r}, \vghp)-n(x, \vek{r}, \vgh)]
\nonumber\\
&\;
+\frac{3}{16\pi} \The \int \id^2\vghp \left[2-4\mu_{\rm sc}-6\mu^2_{\rm sc}+4\mu^3_{\rm sc}\right] n(x, \vek{r}, \vghp)
\\
\nonumber
&\quad
+\frac{3}{16\pi} \frac{\The}{x^2}\frac{\partial}{\partial x} x^4\frac{\partial}{\partial x} 
\int \id^2\vghp (1+\musc^2)(1-\musc) \, n(x, \vek{r}, \vghp),
\end{align}
where $\musc=\vgh\cdot\vghp$. Stimulated scattering and electron recoil were neglected, because for SZ clusters $\Tg\ll \Te$.

The first integral in Eq.~\eqref{eq:gen_coll_Te} describes the effect of Thomson scattering (no energy exchange or equivalently scattering of photons by resting electrons), leaving the photon spectrum of the incoming radiation field unaltered. If the incoming photon field is isotropic this term vanishes.

The second integral in Eq.~\eqref{eq:gen_coll_Te} accounts for leading order temperature corrections to the total scattering cross-section. These again do not alter the shape of the photon spectrum (no energy exchange), and vanish for cold/resting electrons. For isotropic incoming photon field this term cancels [Klein-Nishina corrections, leading to contributions $\simeq \The (h\nu/\me c^2)$ are neglected here], but for the radiation dipole through octupole a first-order temperature correction arises:
although for resting electrons the Thomson cross-section only couples to the monopole and quadrupole anisotropy of the incoming photon field, once electrons are moving, due to light aberration \citep[see e.g.,][for related discussion]{Chluba2011ab}, also photons from the radiation dipole and octupole scatter into the line of sight at $\mathcal{O}(\The)$. Higher order multipoles couple at order $\mathcal{O}(\The^{\ell-2})$. 

The last integral in Eq.~\eqref{eq:gen_coll_Te} takes the energy exchange between electrons and photons into account. It depends only on the radiation monopole through octupole.  If the incoming radiation monopole through octupole have identical energy distribution, then their spectra are all altered in the same way by the scattering event. The difference is captured by an overall, frequency independent scattering efficiency factor that depends on the multipole.
This is the case relevant to the second scattering SZ correction, for which all multipoles of the singly-scattered radiation field have a spectrum $\propto Y_0(x)$, which represents the well-known thSZ spectrum \citep{Zeldovich1969} defined in Eq.~\eqref{eq:SZ_formula}.

\subsection{Rewriting the collision term using Legendre polynomials}
The collision term can be further simplified by expanding the radiation field in terms of Legendre polynomials, $P_\ell(x)$:
\beal
\label{eq:Leg_n}
n(x, \vek{r}, \musc)&=\sum_{\ell} P_\ell(\musc) \, n_\ell(x, \vek{r})
\nonumber\\
n_\ell(x, \vek{r})&=\frac{2\ell+1}{4\pi} \int  \id^2\vghp P_\ell(\musc) \,n(x, \vek{r}, \vghp).
\end{align}
Here, $n(x, \vek{r}, \musc)$ denotes the azimuthally averaged occupation number of the photon field along the propagation direction $\vgh$. The Legendre coefficients can also be obtained from a spherical harmonic expansion of the radiation field with respect to a general system, $n(x, \vek{r}, \vghp)=\sum_{\ell m} Y_{\ell m}(\vghp) n_{\ell m}(x, \vek{r})$, using the identity $n_\ell(x, \vek{r})\equiv \sum_{m} Y_{\ell m}(\vgh) \,n_{\ell m}(x, \vek{r})$. Carrying out the integrals in Eq.~\eqref{eq:gen_coll_Te}, we  find \citep[compare Eq. (C15) and (C19) of][]{Chluba2012}:
\beal
\label{eq:final_coll_Te}
\mathcal{C}[n] &\approx n_0(x, \vek{r})+\frac{1}{10} n_2(x, \vek{r}) - n(x, \vek{r}, \vgh)
\nonumber\\
&\!\!\!
-\The\left[\frac{2}{5}n_1(x, \vek{r})+\frac{3}{5} n_2(x, \vek{r}) - \frac{6}{35}n_3(x, \vek{r}) \right]
\\
\nonumber
&\!\!\!
+\frac{\The}{x^2}\frac{\partial}{\partial x} x^4\frac{\partial}{\partial x} 
\left[ n_0(x, \vek{r})-\frac{2}{5} n_1(x, \vek{r})+\frac{1}{10}  n_2(x, \vek{r}) - \frac{3}{70} n_3(x, \vek{r}) \right].
\end{align}
This expression captures all the relevant physics of the collision term, Eq.~\eqref{eq:gen_coll_Te}. Like Eq.~\eqref{eq:Coll_monopole}, it can be directly obtained by computing the first and second moments of the frequency shift over the scattering kernel when accounting for the full angular dependence of the incoming photon field (see \citet{Chluba2012} and CNSN). Only multipoles with $\ell\leq 3$ really matter for the scattering problem, greatly simplifying the next steps of the calculation. This expressions also clearly shows that only the azimuthally symmetric part of the incoming photon distribution with respect to $\vgh$ really matters, a property that due to symmetries of the scattering problem also holds at higher temperature (see CD13).

\subsection{Born approximation for multiple scattering}
In the optically thin limit ($\tau\ll 1$), relevant for galaxy clusters, repeated Compton scattering only introduces a small correction to the radiation field. To solve Eq.~\eqref{eq:Boltzmann}, we can thus use a perturbative Ansatz (a `Born-series'),
$n(x,\vek{r}, \vgh) \approx n^{(0)} + \epsilon \, n^{(1)} + \epsilon^2 n^{(2)} + ... + \epsilon^k n^{(k)}$, yielding the iteration scheme
\beal
\label{eq:Boltzmann_pert}
\frac{\partial n^{(k)}(x,\vek{r}, \vgh)}{\partial \tau}&\approx \mathcal{C}[n^{(k-1)}(x,\vek{r}, \vgh)].
\end{align}
The Compton collision term is linear in $n(x,\vek{r}, \vgh)$ since induced scattering terms cancel for $\Tg\ll \Te$. Without scattering nothing changes ($\partial_\tau n^{(0)}(x,\vek{r}, \vgh) = 0$), and the first scattering signal is given by $\partial_\tau n^{(1)}(x,\vek{r}, \vgh) = \mathcal{C}[n^{(0)}(x,\vek{r}, \vgh)]$.
For higher order scatterings, the local anisotropy of the radiation field becomes important, even if the unscattered radiation field was isotropic. 
Since the optical depth of typical clusters is $\tau \lesssim 0.01$, we can restrict ourselves to the first and second scattering effect.

\section{Multiple scattering SZ contributions at lowest order in electron temperature}
\label{sec:multiple_scatt_lowest}

\subsection{SZ signal after the first scattering}
Assuming that the incoming, unscattered radiation field is given by the isotropic CMB blackbody, $n^{(0)}(x,
\vek{r}, \vgh)= \nbb(x)\equiv 1/[\expf{x}-1]$, the only non-vanishing incoming-radiation multipole is the monopole
($\ell=0$).  The collision term in this case simplifies to the well-known form \citep[see][]{Zeldovich1969},
\beal
\label{eq:Coll_monopole}
\mathcal{C}_0[n]\approx \frac{\The}{x^2}\frac{\partial}{\partial x} x^4\frac{\partial}{\partial x} n(x)
=\The \left[4 x\frac{\partial}{\partial x} +x^2\frac{\partial^2}{\partial x^2} \right]  n(x),
\end{align}
for an isotropic radiation field.  The coefficients, $4\The$ and
$\The$, of the derivative terms are simply the first and second
moments of the scattered photon's relative frequency shift,
$\Delta_\nu \equiv (\nu'-\nu)/\nu$, over the scattering kernel
\citep[e.g., see][]{Sazonov2000}.

This then leads to the well-known thSZ effect \citep{Zeldovich1969}, with spectral shape,
\beal
\label{eq:SZ_formula}
\Delta I_\nu/I_{\rm o} \approx y\, \frac{x^4\expf{x}}{(\expf{x}-1)^2}
\left[x\,\frac{\expf{x}+1}{\expf{x}-1}-4\right] \equiv y \, x^3 Y_0(x).
\end{align}
Here, the Compton $y$ parameter, $y(\vgh)= \int (k\Te/\me c^2)\,  \sigT\Ne  \id l$, for a distant observer is computed along the line of sight through the ICM, and $I_{\rm o}=(2h/c^2)(kT_0/h)^3\approx 270\,\MJy\,\sr^{-1}$. We also assumed that the cluster is at rest in the CMB rest frame and internal motions are negligible (no kSZ).

\begin{figure}
\centering
\includegraphics[width=\columnwidth]{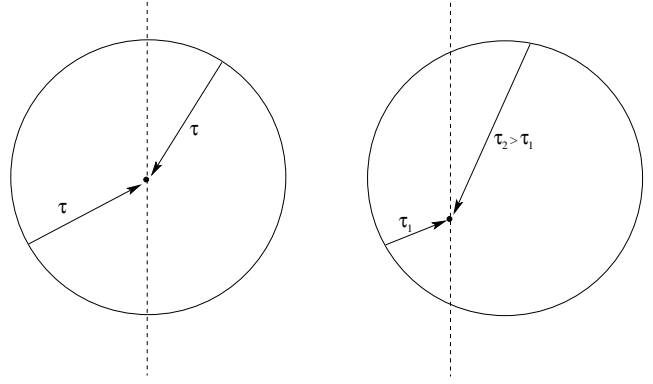}
\caption{Optical depth anisotropy of the local radiation field introduced by the first scattering of CMB photons for a constant density, isothermal sphere of free electrons. The dashed lines indicate the lines-of-sight for the two cases, and the dot marks the locus of the second scattering event. At the center (left panel), the total optical depth in any direction is the same because the total path through the medium is identical. Observing the singly-scattered radiation field around an off center position (right panel) on the other hand shows anisotropies because of optical depth variations.}
\label{fig:spherical_cloud}
\end{figure}
\subsection{Second scattering correction}

\subsubsection{Simple illustration for the scattering-induced anisotropy}
While the derivation given in this section applies to general cluster geometry, to illustrate the physics of the multiple scattering problem, let us consider a spherical, isothermal cloud of free electrons at rest in the CMB rest frame with constant density, $\Ne$, and radius $R$ as a simple example. 
At any location, $\vek{r}$, inside the cluster we can compute the singly-scattered radiation field in different directions using Eq.~\eqref{eq:SZ_formula}, with isothermal $y$ parameter, $y(\vek{r}, \vghp)=\The \tau(\vek{r}, \vghp)$. Here, the optical depth, $\tau(\vek{r}, \vghp)=\int \sigT \Ne \id s$, is computed from $\vek{r}$ in the direction $\vghp$ through the ICM, where $s$ parametrizes the length along photon path.
At the cluster center, the incident radiation field is fully isotropic, even in the limit of infinite number of scatterings; however, due to variations of the scattering optical depth in different directions, after one scattering the local photon field becomes anisotropic elsewhere (see Fig.~\ref{fig:spherical_cloud}). This {\it scattering-induced anisotropy}, generally being sourced by both electron temperature and density variations, has to be taken into account when calculating the second scattering correction to the SZ signal.

\subsubsection{Explicit form of the second scattering SZ correction}
\label{sec:explicit_expression_lowest}
We now give the explicit expression of the second scattering correction to the thSZ effect. After the first scattering, the correction to the radiation field at any position $\vek{r}$ inside the cluster is given by 
\beal
\label{eq:SZ_first_general}
n^{(1)}(x, \vek{r}, \vgh')\approx y(\vek{r}, \vgh') Y_0(x), 
\end{align}
where the spatial dependence is introduced only by the variation of the $y$ parameter in different directions $\vgh'$, with the scatterer centered at position $\vek{r}$. 
This expression shows that for the second scattering SZ signal, spectral and spatial parts of the radiation field separate. Since all multipoles of the singly-scattered radiation have the same spectrum, they also have the same after the second scattering. When including higher order temperature corrections and variations of the electron temperature along different lines-of-sight, this separation becomes slightly more complicated and the spectra of each multipole differ after the second scattering (see CD13).

Inserting Eq.~\eqref{eq:SZ_first_general} into Eq.~\eqref{eq:final_coll_Te}, and carrying out the line of sight average, we find the second scattering correction to the thSZ signal:
\bsub
\label{eq:second_lowest_DI}
\beal
\label{eq:second_lowest_DI_a}
\Delta I^{(2)}(x, \vgh)&=\Delta I^{(2), \rm T}(x, \vgh)+\Delta I^{(2), \sigma}(x, \vgh)+\Delta I^{(2), \rm E}(x, \vgh)
\\[1mm]
\label{eq:second_lowest_DI_b}
\Delta I^{(2), \rm T}/I_{\rm o}
&\approx x^3 Y_0(x) \left[ \left< y_0\right> + \frac{1}{10}\left< y_2\right> - \left< y(\vek{r},\vgh)\right> \right]
\\
\Delta I^{(2), \sigma}/I_{\rm o}
&\approx
x^3 Y_0(x) 
\left[-\frac{2}{5}\left<\The  y_1\right>-\frac{3}{5}\left<\The  y_2\right>+\frac{6}{35}\left<\The  y_3\right>\right]
\nonumber\\
\Delta I^{(2), \rm E}/I_{\rm o}
&\approx
x^3 Y^{\ast}_0(x)  
\left[\left<\The  y_0\right>
-\frac{2}{5}\left<\The  y_1\right>
+\frac{1}{10}\left<\The  y_2\right>
-\frac{3}{70}\left<\The  y_3\right>\right]
\nonumber\\[1mm]
\label{eq:Yast}
Y^\ast_0(x)&=x^{-2}\partial_x x^4 \partial_x Y_0(x)
\nonumber\\
&=120\,G(x)+34\,Y_0(x) + 12x\,G(x)\,Y_0(x)
\\ \nonumber
&\qquad\quad 
+2x\,G(x)^2\left[x \sinh(x)-12\cosh(x)\right]
\nonumber\\
\label{eq:second_lowest_DI_c}
 \left<X \right>&=\int X(\vek{r}) \sigT \Ne(\vek{r}) \id z.
\end{align}
\esub
with $G(x)=x\expf{x}/[\expf{x}-1]^2$.
Here, $z$ parametrizes the photon path along the line of sight $\vgh$. We also introduced the Legendre coefficients, $ y_\ell$, of $y(\vek{r}, \vgh')$, which are defined like in Eq.~\eqref{eq:Leg_n}. 

Equation~\eqref{eq:second_lowest_DI} describes the full second scattering correction at lowest order in the electron temperature. The first two lines in Eq.~\eqref{eq:second_lowest_DI_b} provide corrections, $\Delta I^{(2), \rm T}$ and $\Delta I^{(2), \sigma}$, to the amplitude of the usual thSZ distortion, Eq.~\eqref{eq:SZ_formula}. The second is smaller by a factor $\The \ll 1$ compared with the first, and so we neglect it below.  The third provides the new frequency dependence, proportional to $Y^\ast_0(x)$, induced in the second scattering. The results indicate that only radiation multipoles up to $\ell=3$ (the octupole) are required to describe the radiation field obtained in second scattering.

\begin{figure}
\centering
\includegraphics[width=\columnwidth]{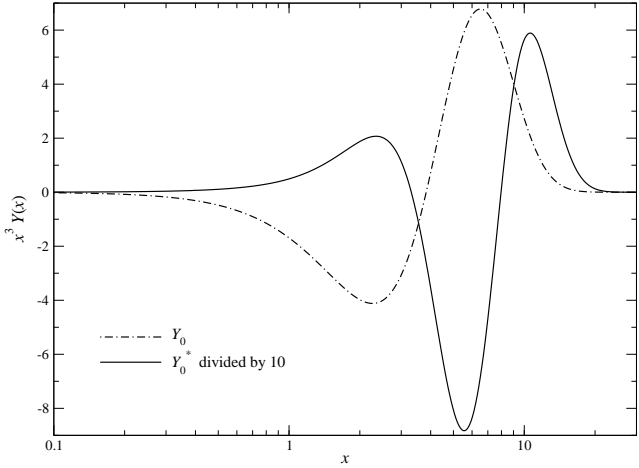}
\caption{Spectral shape of the lowest order first and second scattering contributions to the thSZ signal. Note that we scaled $Y^\ast_0(x)=x^{-2}\partial_x x^4 \partial_x Y_0$ to make it more comparable in amplitude to $Y_0$.}
\label{fig:lowest_order_Y_functions}
\end{figure}

Since the energy-exchange correction, $\Delta I^{(2), \rm E}$, has a frequency dependence that differs from that of the usual lowest order thSZ effect, measurement of its amplitude can provide information about anisotropies in the electron temperature and density profiles.  As shown in Fig. 2, the function $Y^\ast_0(x)$ numerically reaches values $\simeq 10$ times larger than the maximum of $Y_0(x)$.  This makes this second-scattering correction roughly an order of magnitude larger than would be guessed from its $\The \simeq 10^{-2}$ suppression relative to the lowest order SZ signal.

To calculate the observable emergent specific intensity, we only need to compute the line of sight averages of the $y$-parameter multipoles, weighted by the Thomson optical depth, $\propto \left< y_\ell\right>$, and the $y$ parameter itself, $\propto \left<\The  y_\ell\right>$. These {\it moments} and 
\beal
\label{eq:identity_start}
\left<y(\vek{r},\vgh)\right>&=\int_{-\infty}^\infty \left(\int_z^\infty \The(\vek{r}') \sigT \Ne(\vek{r}') \id z' \right) \sigT \Ne(\vek{r}) \id z
\nonumber\\
&=\int_0^{\tau(\vgh)} \left(\int^{\tau(\vgh)}_{\tau''} \The(\vek{r}') \id \tau' \right) \id \tau''
\end{align}
can be obtained directly from cluster simulations. This greatly simplifies the computation, since spatial and spectral terms have been separated by our method. Below we give some simple examples assuming spherical symmetry.

\subsection{Second scattering SZ correction in the isotropic scattering approximation (ISA)}
\label{sec:iso_approx}
In the ISA \citep[e.g.,][]{Itoh2001, Colafrancesco2003, Shimon2004}, the singly-scattered photon field is assumed to be isotropic and the same as in the direction along the line of sight defined by $\vgh$. Explicitly this means $y(\vek{r}, \vghp) \approx y(\vek{r}, \vgh)$, where $\vghp$ is the direction of the incoming photon. Therefore, none of the anisotropies are taken into account and even the variation of the local monopole is neglected, although generally $\int y(\vek{r}, \vghp)\id^2\vghp \neq y(\vek{r}, \vgh)$.
At lowest order in $\The$, with Eq.~\eqref{eq:second_lowest_DI} we therefore have
\beal
\label{eq:dist_second_iso_approx}
\Delta  I^{(2), \rm iso}(x, \vgh)\approx \Delta  I^{(2), \rm E, \rm iso}(x, \vgh) 
&\approx
I_{\rm o} x^3 Y^\ast_0(x) \, \frac{y(\vgh)^2}{2},
\end{align}
where we used the identity $\left<\The y(\vek{r}, \vgh)\right>\equiv y(\vgh)^2 / 2$, which can be deduced using Eq.~\eqref{eq:identity_start}.
This expression neglects geometric {\it form factors}. In addition, no Thomson correction, $\Delta  I^{(2), \rm T}(x, \vgh)$, is obtained in the ISA, although it usually is comparable to the energy-exchange correction, $\Delta  I^{(2), \rm E}(x, \vgh)$, in Eq.~\eqref{eq:second_lowest_DI} [see Fig.~\ref{fig:total_corr_lowest}].

\subsection{Total SZ signal with second scattering corrections}
\label{sec:total_signal_def}
The second scattering SZ signal gives rise to new observables that depend on the geometry of the cluster. 
The required line of sight moments scale like $\left< y_l\right>\propto \tau(\vgh) y(\vgh) / 2$, $\left<\The  y_l \right>\propto y^2(\vgh) / 2$ and $\left<y(\vek{r},\vgh)\right>\propto \tau(\vgh) y(\vgh) / 2$. 
With Eq.~\eqref{eq:SZ_formula} and Eq.~\eqref{eq:second_lowest_DI}, it is therefore useful to rewrite the total SZ signal as
\beal
\label{eq:total_SZ_signal_r}
\Delta  I (x, \vgh)/I_{\rm o} & 
\approx 
x^3 Y_0(x) \, y(\vgh) \left( 1+  \frac{\tau(\vgh)}{2}\,\gamma_{\rm T}(\vgh)\right)
+x^3 Y^\ast_0(x) \, \gamma_{\rm E}(\vgh) \, \frac{y^2(\vgh)}{2} 
\nonumber \\
\gamma_{\rm T}(\vgh)&=\frac{2}{y(\vgh)\tau(\vgh)}\left[ \left< y_0\right> + \frac{\left< y_2\right>}{10} - \left<y(\vek{r},\vgh)\right> \right]
\\
\gamma_{\rm E}(\vgh)&=\frac{2}{y^2(\vgh)}\left[ \left<\The  y_0\right>-\frac{2}{5}  \left< \The  y_1\right>
+ \frac{\left< \The  y_2\right>}{10} 
-\frac{3}{70} \left< \The  y_3\right> \right].
\nonumber
\end{align}
The correction due to Thomson scattering is determined by $\gamma_{\rm T}(\vgh)$, while the energy-exchange correction depends on $\gamma_{\rm E}(\vgh)$. For the ISA, we have $\gamma_{\rm T}(\vgh)=0$ and $\gamma_{\rm E}(\vgh)=1$.

Since the total SZ signal depends on two independent spectral functions, $Y_0(x)$ and $Y^\ast_0(x)$ [see Fig.~\ref{fig:lowest_order_Y_functions}], from the observational point of view it in principle is possible to directly measure $A(\vgh)=y(\vgh) \left[ 1+  \tau(\vgh)\,\gamma_{\rm T}(\vgh)/2\right]$ and $B(\vgh)=\gamma_{\rm E}(\vgh) \,y^2(\vgh)/2$ for different lines-of-sight. 
These observables depend on both the temperature and density profiles of the electron distribution, and thus allow probing properties of the cluster atmosphere. 

\section{Spherically symmetric systems}
\label{sec:spherical}
For spherically symmetric systems, the computations of the required $y$-parameter moments can be further simplified. Due to the symmetry of the problem, the $y$ parameter in any direction $\vghp$ around location $\vek{r}$ is given by 
\beal
\label{eq:gen_y_sphere}
y(\vek{r}, \vghp)\equiv y(r, \mu_r) = \int_0^\infty \sigT \,\The(r^\ast) \Ne(r^\ast) \id s
\end{align}
with $r^\ast(s, \mu_r)=\sqrt{r^2+2r s \mu_r + s^2}$ and $\mu_r=\vek{\hat r}\cdot \vghp$, where $\vek{\hat r}$ defines the radial direction. 
To compute the Legendre coefficients of $y(\vek{r}, \vghp)$, we can use the addition theorem for spherical harmonics to simplify the calculation. This gives
\beal
\label{eq:Leg_y}
 y_\ell(\vek{r})&=\frac{2\ell+1}{4\pi} \int  \id^2\vghp \,P_\ell(\vgh\cdot \vghp) \,y(\vek{r}, \vghp)
\nonumber\\
&= \sum_m \int  \id^2\vghp \,Y_{\ell m}(\vgh)Y^\ast_{\ell m}(\vghp) \,y(r, \mu_r)
\nonumber\\
&= \frac{2\ell+1}{2}  \int  \id \mu_r \,P_\ell(\vek{\hat r}\cdot\vgh)\,P_\ell(\mu_r) \,y(r, \mu_r)
\nonumber\\
&=P_\ell(z/r) \,  y^\ast_\ell(r) \equiv  y_\ell(b, z),
\end{align}
where we introduced the impact parameter $b$ of the line of sight $\vgh$ from the cluster center at the position $z$, so that $r=\sqrt{b^2+z^2}$. 
The Legendre coefficient $ y^\ast_\ell(r)\equiv  y_\ell(b=0, z)$ describes the simplest case, i.e., the line of sight through the center, and the factor $P_\ell(z/r)$ provides the geometric transformation to the case $b\neq 0$.
From the practical point of view, this means that for spherically symmetric systems we can compute the required line of sight moments $\left< y_\ell\right>$ and $\left<\The  y_\ell\right>$ in two independent steps, first giving $ y^\ast_\ell(r)$ and then averaging over $z$. Since $ y_\ell(b, -z)=(-1)^\ell y_\ell(b, z)$, all odd moments vanish after averaging over the line of sight. For general cluster geometry this is not necessarily true but it greatly simplifies the calculation for (quasi) spherically systems.

Performing the required integrals, the parameters $\gamma_{\rm E}(b)$ and $\gamma_{\rm T}(b)$ in
Eq.~\eqref{eq:total_SZ_signal_r}, as a function of impact parameter $b$ become (see Appendix~\ref{app:explicit_spherical} for more details)
\bsub
\label{eq:gammaEspherical}
\beal
     \gamma_{\rm E}(b) &= \frac{\sigT}{\me c^2}
     \frac{2}{[y_{\rm obs}(b)]^2} \sum_{\ell=0,2} \kappa_\ell
      \\ \nonumber
     & \qquad \times
     \int^{\sqrt{R^2-b^2}}_0 \!\!\id z \, P_{\rm e}(r)
     P_\ell\left(\frac{z}{r}\right) \int^1_0 \!\id\mu
     \, P_\ell(\mu) \, y_{\rm obs}\left( r \sqrt{1-\mu^2}
     \right), \\ \nonumber
     \gamma_{\rm T}(b) & = \frac{2\sigT}{y_{\rm obs}(b) \tau_{\rm
     obs}(b)} \left\{ \sum_{\ell=0,2} \kappa_\ell
      \right. \nonumber \\
     &\qquad  \times 
     \int^{\sqrt{R^2-b^2}}_0 \!\!\id z \, N_{\rm e}(r)
     P_\ell\left(\frac{z}{r}\right) \int^1_0 \!\id \mu\,
     P_\ell(\mu) \,y_{\rm obs}\left( r \sqrt{1-\mu^2}
     \right)  \nonumber\\
     &\qquad\qquad  \left. - 2 \int^{\sqrt{R^2-b^2}}_0\! \id z\,
     \Ne(r) \, y\left(r,\frac{z}{r}\right)
     \right\},
\end{align}
\esub
in terms of $y_{\rm obs}(b)$, the $y$ parameter observed along
a line of sight that passes an impact parameter $b$ from the
center.  In these expressions, $r=\sqrt{z^2+b^2}$, and
$\kappa_0=1$ and $\kappa_2=1/2$. We also assumed that the cluster extends only to a maximal radius $R$ and introduced the pressure profile $P_{\rm e}(r)=k\Te(r) \Ne(r)$. To simplify the computation, the innermost integrals over $\id\mu$ can be tabulated as a function of $r$.

\begin{figure}
\centering
\includegraphics[width=\columnwidth]{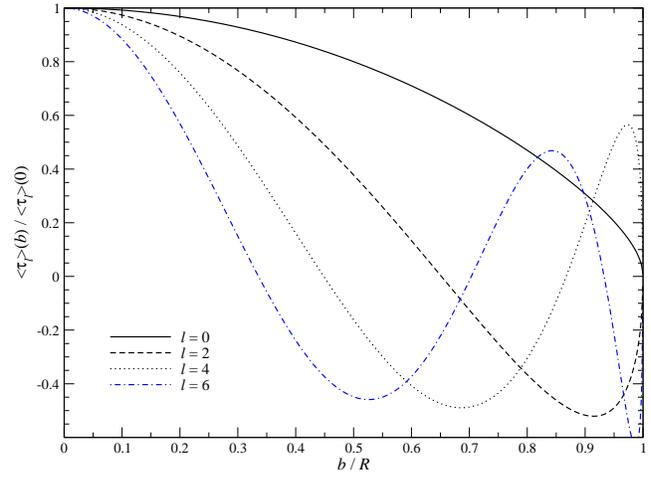}
\caption{Weighted average optical depth multipoles for an isothermal, constant density sphere with radius $R$ as a function of impact parameter. All curves were normalized to their central value $\left<\tau_{\ell}\right>(0)=\left<\tau^\ast_{\ell}\right>$, with $\tau_{\rm c}=2R\Ne \sigT$, $\left<\tau^\ast_{0}\right>=0.87\,\tau_{\rm c}^2/2$,  $\left<\tau^\ast_{2}\right>=0.15\,\tau_{\rm c}^2/2$, $\left<\tau^\ast_{4}\right>=-0.016\,\tau_{\rm c}^2/2$ and $\left<\tau^\ast_{6}\right>=0.0045\,\tau_{\rm c}^2/2$ according to Eq.~\eqref{eq:r_l_transforms}. Odd moments vanish.}
\label{fig:off_center}
\end{figure}

\begin{figure}
\centering
\includegraphics[width=\columnwidth]{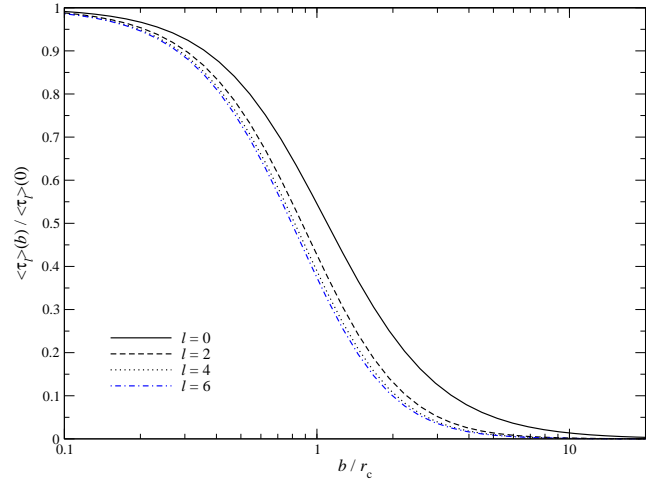}
\caption{Weighted average optical depth multipoles for isothermal $\beta$-profile ($\beta=2/3$) with core radius, $r_{\rm c}$, as a function of impact parameter. All curves where normalized to their central value $\left<\tau_{\ell}\right>(0)=\left<\tau^\ast_{\ell}\right>$, with $\tau_{\rm c}\approx 3.141\,N_{\rm e} \sigT r_{\rm c}$, $\left< \tau^\ast_{0}\right>=0.69\,\tau_{\rm c}^2/2$,  $\left<\tau^\ast_{2}\right>=0.14\,\tau_{\rm c}^2/2$, $\left<\tau^\ast_{4}\right>=0.054\,\tau_{\rm c}^2/2$ and $\left<\tau^\ast_{6}\right>=0.028\,\tau_{\rm c}^2/2$.
Odd moments vanish by symmetry.
}
\label{fig:off_center_iso_beta}
\end{figure}

\subsection{Optical depth moments for illustrative cases}
\label{sec:sphericalopticaldepth}
The second scattering SZ signal, Eq.~\eqref{eq:total_SZ_signal_r}, directly depends on the {\it geometry} of the cluster atmosphere. General cases can be studied using cluster simulations, but to illustrate the main effects we consider an isothermal, constant density sphere and an isothermal $\beta$-model (see Appendix~\ref{app:tau_l_simple_cases} for definitions). Since for this case $\left< y_\ell\right>=\The \left<\tau_\ell\right>$, $\left<\The  y_\ell\right>=\The^2 \left<\tau_\ell\right>$ and $\left<y(\vek{r}, \vgh)\right>=\The \tau(\vgh)^2 / 2$, we only have to compute line of sight moments of the optical depth field, $\left< \tau_\ell\right>$. Some details of the calculation for $b=0$ are given in Appendix~\ref{app:tau_l_simple_cases}, and the general case is obtained by averaging $ y_\ell(b, z)\equiv P_\ell(z/r) \,  y^\ast_\ell(r)$ along the line of sight.

The dependence of $\left< \tau_\ell\right>$ on impact parameter $b$ for the two cases is illustrated in Fig.~\ref{fig:off_center} and \ref{fig:off_center_iso_beta}. For the constant density sphere, the moments show significant variation with $b$. They drop rapidly with $\ell$, and in all cases the largest values of $\left< \tau_\ell\right>$ are found for the line of sight directly through the center. 
This behavior depends critically on the geometry and density profile of the cloud, and for an isothermal $\beta$-profile the situations is already very different. In contrast to the constant density sphere, the profiles $\left< \tau_\ell\right>$ are monotonic, with higher order terms decreasing more gradually. It is evident that the amplitude and spatial morphology of the second scattering SZ correction differs in the two cases. Consequently, the second scattering correction directly probes the geometry of the scattering medium, an effect that is not captured by the ISA.
%

\subsubsection{Total second scattering signal and comparison to the ISA}
\label{sec:total_corr_lowest_order}
\begin{figure}
\centering
\includegraphics[width=\columnwidth]{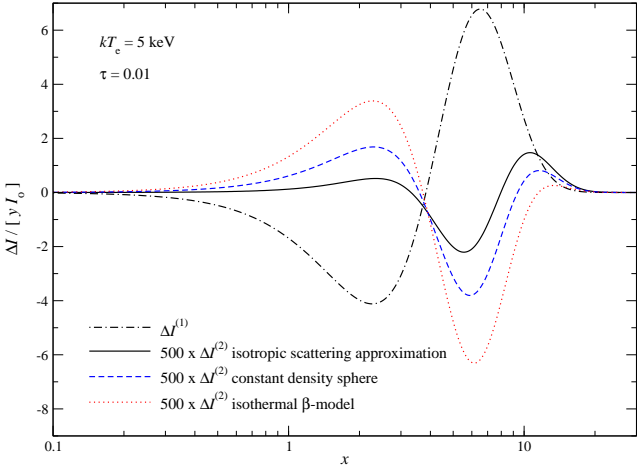}
\caption{Total second scattering correction for the constant density, isothermal sphere and the isothermal $\beta$-model at zero impact parameter. For comparison, we show the single-scattering thSZ effect (the same in all cases), and the second scattering signal obtained with the ISA.}
\label{fig:total_corr_lowest}
\end{figure}

We now compute the total correction caused by second scattering terms. In Fig.~\ref{fig:total_corr_lowest} we show a comparison of the signal for the constant density, isothermal sphere and the isothermal $\beta$-model. The overall signal is very small, in both cases reaching no more than $\simeq 0.2\%$ of $\Delta I^{(1)}$.
However, as Fig.~\ref{fig:total_corr_lowest} illustrates, the second scattering signal directly depends on the geometry of the scattering electron distribution, an effect that is absent in the ISA. 

Comparing the different contributions in Eq.~\eqref{eq:total_SZ_signal_r} also shows that the dominant correction is caused in the Thomson limit, $\Delta I^{(2), \rm T}$. For the isothermal sphere and impact parameter $b=0$, we find $\Delta  I^{(2), \rm T} (x)\approx -0.06 \,\tau_{\rm c} \, \Delta  I^{(1)} (x)$. For some clusters one encounters central optical depth $\tau_{\rm c}\simeq 0.01$, so that this effect gives rise to a correction $\simeq -0.06\%$ relative to the singly-scattered SZ signal.
For an isothermal $\beta$-profile ($\beta=2/3$), we obtain a slightly larger effect, giving $\Delta  I^{(2), \rm T} (x)\approx -0.15 \,\tau_{\rm c} \, \Delta  I^{(1)} (x)$ or a correction $\simeq -0.15\%$ relative to $\Delta  I^{(1)} (x)$ for $\tau_{\rm c} =0.01$.
Comparing the energy-exchange correction with the ISA, we find $\Delta  I^{(2), \rm E, \rm iso}\simeq 1.12  \Delta  I^{(2),  \rm E}$ and $\Delta  I^{(2),  \rm E, \rm iso}\simeq 1.4  \Delta  I^{(2),  \rm E}$ for the isothermal sphere and $\beta$-model, respectively. Hence, this part of the correction is {\it overestimated} when using the ISA. 
The total correction ($\Delta I^{(2), \rm T}+\Delta  I^{(2),  \rm E}$) for the isothermal $\beta$-model is $\simeq 3-4$ times larger than obtained with the ISA.

\subsubsection{Effect on the crossover frequency}
\label{sec:crossover}
Because the frequency dependence of $\Delta  I^{(2),  \rm E}$ differs from $\Delta  I^{(1)}$, the position of the crossover frequency ($\equiv$ null of the SZ signal at $\nu\simeq 217\,\GHz$ or $\xc\simeq 3.83$) is affected by the second scattering contribution. From Eq.~\eqref{eq:total_SZ_signal_r}, we find
\beal
\label{eq:crossover}
\Delta \xc &\approx 3.1 \, y(\vgh) \,\gamma_{\rm E}(\vgh),
\end{align}
which for the isothermal $\beta$-model ($\gamma_{\rm E} \simeq 0.71$ at $b=0$) with total $y$ parameter $y\simeq 10^{-4}$ means $\Delta \xc \simeq \pot{2.2}{-4}$. For the constant density sphere ($\gamma_{\rm E}\simeq 0.89$ at $b=0$), we find $\Delta \xc \simeq \pot{2.7}{-4}$, while the ISA gives $\Delta \xc \simeq \pot{3.1}{-4}$. 

For comparison, at $\Te\lesssim 30\,\keV$ the shift in the position of the crossover frequency caused by higher order temperature corrections is $\Delta \xc\simeq 0.0421\hat{T}_{\rm e}-\pot{3.29}{-4}\hat{T}_{\rm e}^2$ with $\hat{T}_{\rm e}=\Te/5\keV$. For $\The=0.01 (\equiv \hat{T}_{\rm e}\simeq 1)$, this means $\Delta \xc\simeq 0.0418$, which is roughly two orders of magnitudes larger. Similarly, the shift introduced by line of sight variations of the electron temperature is roughly given by $\Delta \xc \simeq 0.042 \hat{T}_{\rm e} [1-0.069 \hat{T}_{\rm e} ] \omega^{(1)}$, where $\omega^{(1)}$ parametrizes the line of sight temperature variance (see CSNN). For $\The=0.01$ and typical temperature-dispersion $\omega^{(1)}\simeq 0.1$ this means $\Delta\xc \simeq \pot{3.9}{-3}$, which is about one order of magnitude larger than the effect caused by multiple scattering. Note that at lowest order in $\Te$, Eq.~\eqref{eq:crossover} agrees with the expressions given by \citet{Dolgov2001} when setting $\gamma_{\rm E}\equiv 1$, i.e., ignoring scattering-induced anisotropies of the radiation.

\begin{figure}
\centering
\includegraphics[width=\columnwidth]{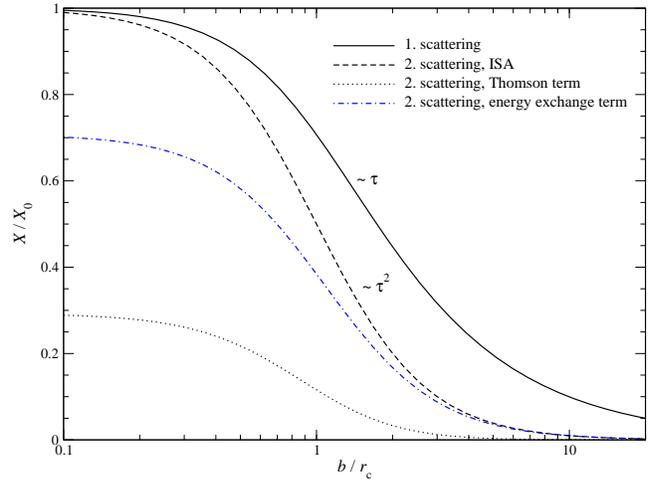}
\caption{Spatial dependence of the different contributions to the second scattering signal for an isothermal $\beta$-profile with $\beta=2/3$. To represent the first scattering we show $y=\tau \The$, renormalized by its central value. For the ISA, we presented $\kappa^{\rm iso}_{\rm E}=(\tau\The)^2 /2$ again renormalizing by its central value.
The Thomson scattering term is $\kappa_{\rm T} =\gamma_{\rm T}\,(\The \tau^2/2)= \The \left[\left<\tau_0\right>+\left<\tau_2\right>/10 - \tau^2/2\right]$ and was normalized by $X_0=-\The \tau^2/2$ at $b=0$.
Finally, the energy-exchange term is $\kappa_{\rm E}=\gamma_{\rm E}\,(\The^2 \tau^2/2)=\The^2 \left[\left<\tau_0\right>+\left<\tau_2\right>/10\right]$ and was normalized by $X_0= (\The \tau)^2/2$ at $b=0$.}
\label{fig:spatial}
\end{figure}

\subsubsection{Spatial dependence of the total second scattering signal}
\label{sec:total_corr_spatial}
One interesting aspects of the second scattering correction is that it introduces {\it both} a new spatial and spectral dependence. At lowest order in temperature, the singly-scattered signal has a spectrum $\simeq x^3 Y_0(x)$ and spatial variation $y\simeq\tau \The$. In the ISA, the second scattering contribution [cf. Eq.~\eqref{eq:dist_second_iso_approx}] has spectrum $\simeq x^3 Y^{\ast}_0(x)$ and spatial dependence $\kappa^{\rm iso}_{\rm E}\simeq(\tau\The)^2 /2$.
On the other hand, when including the anisotropy of the singly-scattered radiation field, two new contributions arise, each with individual spatial and spectral dependence [see Eq.~\eqref{eq:second_lowest_DI}]. The Thomson correction term scales like $\kappa_{\rm T}=\gamma_{\rm T}\,(\The \tau^2/2)\simeq \The \left[\left< \tau_0\right>+\left<\tau_2\right>/10 - \tau^2/2\right]$ with a spectrum $\simeq x^3 Y_0(x)$, while the energy-exchange term has spectrum $\simeq x^3 Y^\ast_0(x)$ and spatial dependence $\kappa_{\rm E}=\gamma_{\rm E}\,(\The^2 \tau^2/2)\simeq\The^2 \left[\left<\tau_0\right>+\left<\tau_2\right>/10\right]$ ($\left<\tau_1\right>=\left<\tau_3\right>=0$ by symmetry).

In Fig.~\ref{fig:spatial} we illustrate the spatial dependence of different terms for an isothermal $\beta$-profile with $\beta=2/3$. 
The second scattering correction is most important at $b\simeq 0$, exhibiting a $\Ne^2$-scaling like X-rays. Thus, the detectability of the second scattering signal may be enhanced by cross-correlating with X-ray maps, as the X-ray emission also scales with $\Ne^2$. Similarly, part of the SZ polarization signature \citep[e.g.,][]{Sunyaev1980, Sazonov1999, Lavaux2004, Shimon2006} caused by the second scattering should correlate spatially with the SZ intensity signal. 

\subsection{Angle-averaged SZ signal}
\label{sec:averaged_SZ}
For spherical systems with finite radius $R$, we can also compute the angle-averaged SZ signal, assuming that the cluster is filling a circular beam with radius $R$. It is well known, that the average single-scattering signal depends on the volume integral over the electron pressure profile, $P_{\rm e}(r)=k\Te(r) \Ne(r)$
\beal
\label{eq:av_y}
\bar y &= \frac{1}{\pi R^2} \int b \id b \id \varphi \int \The\left(\sqrt{z^2+b^2}\right) \Ne \left(\sqrt{z^2+b^2}\right) \sigT \id z
\nonumber\\
&=\frac{1}{\pi R^2} \frac{\sigT}{\me c^2} \int P_{\rm e}(r) \id V.
\end{align}
For the second scattering signal, geometric form factors appear, which depend on the cluster atmosphere. This can be seen from
\bsub
\beal
\label{eq:av_yl}
\overline{\left< y_\ell\right>} 
&=
\frac{1}{\pi R^2} \int b \id b \id \varphi \int P_\ell\left(z/r\right) 
 y^\ast_\ell\left(r\right) \Ne \left(r\right) \sigT \id z
\nonumber\\
&=\frac{\sigT}{\pi R^2} \int P_\ell (\mu_r)\,  y^\ast_\ell (r) \Ne (r) \id \varphi \id \mu_r r^2 \id r
\nonumber\\
&=\delta_{\ell 0} \frac{\sigT}{\pi R^2} \int \left(\frac{1}{4\pi} \int \The\left(r^\ast\right) \Ne \left(r^\ast\right) \sigT \id s \id \varphi' \id \mu'_r \right) \Ne (r) \id V
\nonumber\\
&= \frac{\delta_{\ell 0}}{\pi R^2} \frac{\sigT^2}{\me c^2} \int \left(\int \frac{P_{\rm e}\left(|\vek{r}+\vek{r}'|\right)}{4\pi \,{r'}^2} \id V' \right)\Ne (r) \id V
\nonumber\\
&= \frac{\delta_{\ell 0}}{\pi R^2} \frac{\sigT^2}{\me c^2} 
\int \int \frac{P_{\rm e}\left(r'\right)\Ne (r)}{4\pi \,{|\vek{r}-\vek{r}'|}^2} \id V'  \id V
\\[2mm]
\overline{\left<\The y_\ell\right>} 
&=\frac{\delta_{\ell 0}}{\pi R^2} \left(\frac{\sigT}{\me c^2}\right)^2 \!\!
\int \int \frac{P_{\rm e}\left(r'\right) P_{\rm e} (r)}{4\pi \,{|\vek{r}-\vek{r}'|}^2} \id V' \id V
\end{align}
\esub
where we used $r^\ast=\sqrt{r^2+2rs\mu'_r + s^2}\equiv |\vek{r}+\vek{r}'|$.  
In the ISA, we have $\overline{\left< y_0\right>}_{\rm ISA}=0$ and $\overline{\left<\The y_0\right>}_{\rm ISA}=\bar y^2/2$. Similar to Eq.~\eqref{eq:total_SZ_signal_r}, it thus makes sense to define the {\it scattering form factors} 
\beal
\label{eq:gamma_av}
\bar \gamma_{\rm T}&=\frac{\overline{\left< y_0\right>}-\overline{\left< y(\vek{r},\vgh)\right>}}{\bar y \bar \tau/2}
\qquad \text{and} \qquad 
\bar \gamma_{\rm E}=\frac{\overline{\left<\The  y_0\right>}}{\bar y^2/2}.
\end{align}
With Eq.~\eqref{eq:SZ_formula} and \eqref{eq:second_lowest_DI}, the total average SZ signal thus reads
\beal
\label{eq:av_Signal}
\overline{\Delta I}(x)/I_{\rm o} &\approx 
x^3 Y_0(x) \, \bar y \left(1 + \bar \gamma_{\rm T} \frac{\bar \tau}{2} \right) 
+ x^3 Y^\ast_0(x) \,\bar \gamma_{\rm E} \frac{\bar y^2}{2}.
\end{align}
The above expressions show that the average SZ signal just depends on the overall geometry of the electron pressure and density distribution, but local anisotropies, that were visible for a resolved cluster, no longer contribute.

The geometric form factors have to be computed on a case-by-case basis. For {\it constant} pressure and electron density profiles, we find $\bar y=(2/3) \, y_{\rm c}$ and $\overline{\left<\The y_0\right>}=(1/4) \, y^2_{\rm c}=(9/16)\, \bar y^2 
\equiv  \overline{\left< y_0\right>}/\The$, where we defined $y_{\rm c}=2\The \Ne \sigT R$ and used $\int \! \!\id V'\! \id V/(4\pi\,|\vek{r}-\vek{r}'|^2)=\pi\,R^4$. This implies $\bar \gamma_{\rm T}+1=\bar \gamma_{\rm E}=1.125 $. 
Assuming that the pressure and electron density profiles scale like $1/r$, we similarly find $\bar y=y_{\rm c}/R$ and $\overline{\left<\The y_0\right>}=[\ln(2)/R^2] \, y^2_{\rm c}=\ln(2)\, \bar y^2 \equiv  \overline{\left< y_0\right>}/\The$, or form factors $\bar \gamma_{\rm T}+1=\bar \gamma_{\rm E}=2\ln(2)\approx 1.4$. 
This shows that depending on the geometry of the cluster profile the ISA scattering approximation only provides a rough approximation for the second scattering signal. In particular it {\it underestimates} the average SZ signal related to $Y^\ast_0(x)$. Although the ISA overestimated the second scattering SZ signal for small impact parameters, at larger radii, which contribute strongly to the averaged signal, it underestimates the signal, explaining this net total effect.

\section{Conclusion}
\label{sec:conclusions}
We analyzed the effect of multiple scattering on the SZ signal demonstrating how this small correction (order $\simeq 0.1\%$) depends on the geometry of the ICM. Previous studies applied the ISA which does not capture the effects discussed here, giving very different results for the second scattering correction (see Fig.~\ref{fig:total_corr_lowest}). The difference has two main sources: (i) due to the anisotropy of the singly-scattered radiation field, the second scattering already introduces a modification in the Thomson limit which is absent in the ISA; (ii) the effective monopole of the radiation field along the line of sight differs from the one used in ISA.

Although, very small in comparison to the singly-scattered signal, our analysis shows that the second scattering SZ signal delivers a new set of observables (see Sect.~\ref{sec:total_signal_def}) which in the future might be useful for 3D cluster-profile reconstruction, allowing us to break geometric degeneracies. 
Since at lowest order in $k\Te/\me c^2$, the second scattering correction for isothermal clusters roughly scales as $\tau^2 \propto \Ne^2$, it introduces an additional spatial dependence of the SZ signal, similar to the one of X-rays. Cross-correlating high-resolution X-ray maps with future SZ maps might thus allow enhancing the detectability of the second scattering signal. Similarly, combining with high-resolution SZ polarization measurements should allow extracting this contribution, potentially opening a path for more detailed studies of the clumpiness of the ICM.

The method developed here can in principle be directly applied to simulated clusters. This can be used to give more realistic estimates for the significance of the second scattering signal. The computation is greatly simplified with our approach, since spatial averages over the ICM can be performed independently of the spectral integrals,
accelerating the calculation.

For the examples discussed here, we assumed spherical symmetry. In this case, all odd moments of the optical depth field vanish; however, in general this is an oversimplification. More realistic cluster atmospheres might exhibit larger differences in the second scattering signal and could thus provide a way to directly probe the asphericity and clumpiness of the ICM along the line of sight. From the observational point of view this is very intriguing, but given the low amplitude of the effect this certainly remains challenging. For our estimates we also assumed isothermality. Variations of the electron temperature cause additional anisotropies in the singly-scattered radiation which might even exceed those introduced by optical depth effects. The second scattering SZ contribution therefore could be used to directly probe the cluster's temperature field. These aspects should, however, be addressed using realistic cluster simulations.

We mention that a similar analysis can be carried out for SZ polarization effects. In particular, the scattering-induced quadrupole anisotropy of the radiation field sources a SZ polarization effect, which is directly related to line of sight average of the Compton $y$ parameter's quadrupole components. In this case, the projections perpendicular to the line of sight are most important (i.e., moments like $\left<y_{2,\pm2}\right>$), defining both amplitude and direction of the polarization signature. Our separation of spatial and spectral dependence should again allow simplifying the calculation of the SZ polarization signature from cluster simulations.
Similarly, our techniques can be applied to other scattering media at small and intermediate optical depth, providing a starting point for efficient numerical schemes, e.g., for cosmological simulations with Lyman-$\alpha$ radiative transfer \citep[e.g.,][]{Maselli2003, Zheng2010}. We plan to explore these possibilities in the future.

\small
\section*{Acknowledgements}
JC thanks Donghui Jeong for stimulating discussions on the problem. The authors are also grateful to Anthony Challinor, Daisuke Nagai and Rashid Sunyaev for useful comments on the manuscript.
This work was supported by the grants DoE SC-0008108 and NASA NNX12AE86G.

\begin{appendix}

\small
\bibliographystyle{mn2e}
\bibliography{Lit}

\small 

\section{Lowest order Fokker-Planck expansion of the collision term}
\label{app:coll_term}
For hot electrons, $\Te \gg T_0$, recoil and stimulated effects do not contribute and the Boltzmann collision term takes the simple form (Sect.~2.1 of CNSN)
\beal
\label{eq:Boltzmann_term}
\mathcal{C}[n]
&=  
\int \frac{f_{\rm e}(p)}{\Ne \sigT} \frac{\id \sigma}{\id \Omega'} 
\left[ n(x', \vek{r}, \vghp) - n(x, \vek{r}, \vgh)\right] \!\id^2\vghp\!\id^3 p.
\end{align}
Here, $f_{\rm e}(p)=\Ne \expf{-\sqrt{1+\eta^2}/\The}/[4\pi \,(\me c)^3 \The K_2(1/\The)]$ is the relativistic Maxwell-Boltzmann distribution with $\eta=p/\me c$. In the required, limit the Compton scattering cross-section reads \citep[e.g., see][]{Jauch1976}
\beal
\label{eq:dsigdO}
\frac{\id\sigma}{\id \Omega'}&\approx\frac{3\sigT}{8\pi}\,\left(\frac{\nu'}{\nu}\right)^2 
\frac{1}{\gamma^2\kappa}
\left[1-\frac{\nu'}{\nu}\frac{\alpha_{\rm sc}}{\gamma^2\kappa^2}
+\frac{1}{2}\left(\frac{\nu'}{\nu}\frac{\alpha_{\rm sc}}{\gamma^2\kappa^2}\right)^2
\right],
\end{align}
with Lorentz factor $\gamma=\sqrt{1+\eta^2}=\eta/\beta$, $\kappa=1-\beta\mu$ and $\alpha_{\rm sc}=1-\mu_{\rm sc}$, where $\mu_{\rm sc}= \vghp\cdot\vgh$ is the cosine of the scattering angle between the incoming and outgoing photon.
Furthermore, $\mu=\vbh\cdot\vgh$ and $\mu'=\vbh\cdot\vghp$ are the direction cosines of the angle between the scattering electron and the incoming and outgoing photon, respectively.
The Thomson scattering cross-section is denoted by $\sigT\approx \pot{6.65}{-25}\,{\rm cm^{2}}$.

Here, we are only interested in contributions at lowest order of the electron temperature, $k\Te/\me c^2\ll 1$. We thus perform a Fokker-Planck expansion of the Boltzmann collision term in the small frequency shift $\Delta_\nu=(\nu'-\nu)/\nu$, replacing $n(x', \vek{r}, \vghp)\approx n(x, \vek{r}, \vghp)+x\partial_x \,n(x, \vek{r}, \vghp)\Delta_\nu+\frac{1}{2}x^2\partial^2_x n(x, \vek{r}, \vghp)\Delta_\nu^2$ in Eq.~\eqref{eq:Boltzmann_term}. This gives
\beal
\label{eq:Boltzmann_expansion}
\mathcal{C}[n]
&\approx \left<n(x, \vek{r}, \vghp)\right>   - n(x, \vek{r}, \vgh) 
+D_x \!\left<n(x, \vek{r}, \vghp)\Delta_\nu \right>
+ \frac{1}{2}\,D_x^2 \!\left<n(x, \vek{r}, \vghp)\Delta_\nu^2 \right>
\nonumber\\
&\left<n(x, \vek{r}, \vghp)\Delta_\nu^k \right>=
\int \frac{f_{\rm e}(p)}{\Ne \sigT} \frac{\id \sigma}{\id \Omega'} 
\,n(x, \vek{r}, \vghp)\Delta_\nu^k \id^2\vghp\!\id^3 p,
\end{align}
with the differential operator $D^m_x=x^m \partial^m_x$. Higher order derivative terms are of higher order in temperature and thus were neglected. 

We now compute the moments of the frequency shift, $\left<n(x, \vek{r}, \vghp)\Delta_\nu^k \right>$. Neglecting the tiny recoil effect we have 
\beal
\label{eq:Delta_nu}
\Delta_\nu&\approx \frac{\beta(\mu'- \mu)}{1-\beta \mu'}\approx \eta(\mu'- \mu)\left[1+\eta\mu'\right]+\mathcal{O}(\eta^3).
\end{align}
Similarly, up to second order of $\eta$, the scattering cross-section takes the form
\beal
\label{eq:Cross_section}
\frac{\id \sigma}{\id \Omega'}&\approx \frac{3\sigT}{16\pi}\left[1+\mu^2_{\rm sc}\right]
+\frac{3\sigT}{16\pi}\left[2\mu'(1-\mu_{\rm sc}+2\mu_{\rm sc}^2)-\mu(1+2\mu_{\rm sc}-\mu_{\rm sc}^2)\right]\eta
\nonumber\\
&\qquad
-\frac{3\sigT}{16\pi}\left[1-2\mu_{\rm sc} +3\mu^2_{\rm sc} - \mu^2(1-\mu_{\rm sc})^2-2\mu'^2(2-4\mu_{\rm sc}+5\mu_{\rm sc}^2)\right.
\nonumber\\
&\qquad\qquad+\Big.4\mu_{\rm sc}(2-\mu_{\rm sc}) \mu \mu' \Big]\eta^2 + \mathcal{O}(\eta^3).
\end{align}
The integrals over electron momenta $p^2\id p=(\me c)^3 \eta^2\id\eta$ give
\beal
\label{eq:relMB_ints}
\int \frac{4\pi}{\Ne}  f_{\rm e}(\eta) \,\eta^2 \id\eta &= 1, \qquad
\int \frac{4\pi}{\Ne}  f_{\rm e}(\eta) \,\eta^3 \id\eta \approx 2\sqrt{2\The/\pi} + \mathcal{O}(\The^{3/2})
\nonumber\\
\int \frac{4\pi}{\Ne}  f_{\rm e}(\eta) \, \eta^4 \id\eta &\approx 3 \The+ \mathcal{O}(\The^2).
\end{align}
We will see that the second integral does not contribute at the end, since the integral over $\mu$ cancels the corresponding terms by symmetry. We now consider the integrals over $\varphi$, defining the azimuthal angle of the scattering electron with respect to $\vgh$. Only $\mu'=\mu_{\rm sc}\mu+\cos(\varphi-\varphi_{\rm sc})(1-\mu^2)^{1/2}(1-\mu_{\rm sc}^2)^{1/2}$ depends on $\varphi$.
At order $\The$ we encounter integrals of the type
\beal
\label{eq:relMB_intstwo}
\int \mu'^0 \id\varphi &= 2\pi, 
\qquad\int \mu' \id\varphi = 2\pi\mu\mu_{\rm sc}
\nonumber\\
\int \mu'^2 \id\varphi &= \pi[1-\mu^2_{\rm sc}-\mu^2(1-3\mu^2_{\rm sc})].
\end{align}
As these expressions show, there is no remaining dependence of the collision term on the azimuthal angle of the scattered photon, $\varphi_{\rm sc}$. We can thus also carry out the integral over $\varphi_{\rm sc}$ for the incoming photon and replace $n(x, \vek{r}, \vghp)\rightarrow \frac{1}{2\pi}\int n(x, \vek{r}, \vghp) \id \varphi_{\rm sc}$.
This gives
\beal
\label{eq:Cross_sectiontwo}
\int \frac{\id \sigma}{\id \Omega'} \frac{\id \varphi\id\varphi_{\rm sc}}{4\pi\sigT}
&\approx \frac{3}{16}\left[1+\mu^2_{\rm sc}\right]
-\frac{3}{16}\left[1+\mu_{\rm sc}^2-4\mu_{\rm sc}^3\right]\eta\,\mu
\nonumber\\
&\quad
+\frac{3}{16}\left[1-2\mu_{\rm sc} +4\mu^3_{\rm sc} -5\mu^4_{\rm sc}\right. 
\nonumber\\
&\qquad- \Big.\mu^2(1-2\mu_{\rm sc}+6\mu_{\rm sc}^2+8\mu_{\rm sc}^3-15\mu^4_{\rm sc}) \Big]\eta^2 
\nonumber\\
\int \frac{\id \sigma}{\id \Omega'}\Delta_\nu \frac{\id \varphi\id\varphi_{\rm sc}}{4\pi\sigT}
&\approx - \frac{3}{16}(1-\mu_{\rm sc})\left[1+\mu^2_{\rm sc}\right]\,\eta\,\mu
\\
&\quad
+\frac{3}{32}(1-\mu_{\rm sc}^2)[3-2\mu_{\rm sc}+5\mu^2_{\rm sc}]\eta^2
\nonumber\\
&\qquad
-\frac{3}{32}(1-\mu_{\rm sc})\left[1+3\mu_{\rm sc} +\mu^2_{\rm sc}+15\mu^3_{\rm sc}\right]\eta^2\mu^2 
\nonumber\\\nonumber
\int \frac{\id \sigma}{\id \Omega'}\Delta_\nu^2 \frac{\id \varphi\id\varphi_{\rm sc}}{4\pi\sigT}
&\approx \frac{3}{32}(1-\mu_{\rm sc})(1+\mu^2_{\rm sc})\left[1+\mu_{\rm sc}+(1-3\mu_{\rm sc})\mu^2\right]\,\eta^2. 
\end{align}
The integrals over $\id \mu$ yield $\int \id \mu=2$, $\int \mu \id \mu=0$, and $\int \mu^2 \id \mu=2/3$, so that finally one is left with 
\beal
\label{eq:moment_Y_lm}
\left< n(x, \vek{r}, \vghp)\right>
&\approx \frac{3}{16\pi}\int 
\id^2\vghp (1+\mu^2_{\rm sc}) \, n(x, \vek{r}, \vghp)
\nonumber\\
&\qquad+\frac{3}{16\pi}\The \int \id^2\vghp
\left[2-4\mu_{\rm sc}-6\mu^2_{\rm sc}+4\mu^3_{\rm sc}\right] n(x, \vek{r}, \vghp)
\nonumber\\
\left<n(x, \vek{r}, \vghp)\Delta_\nu \right>
&\approx \frac{3}{4\pi}\The \int \id^2\vghp
(1+\mu^2_{\rm sc})(1-\mu_{\rm sc})\, n(x, \vek{r}, \vghp)
\nonumber\\
\left<n(x, \vek{r}, \vghp)\Delta_\nu^2 \right>
&\approx \frac{1}{2}\left<n(x, \vek{r}, \vghp)\Delta_\nu \right>.
\end{align}
With Eq.~\eqref{eq:Boltzmann_expansion} and $4D_x + D^2_x = x^{-2}\partial_x x^4 \partial_x$, this then gives Eq.~\eqref{eq:gen_coll_Te}.

\section{Simplifications for spherically symmetric systems}
\label{app:explicit_spherical} 
\begin{figure}
\centering
\includegraphics[width=0.45\columnwidth]{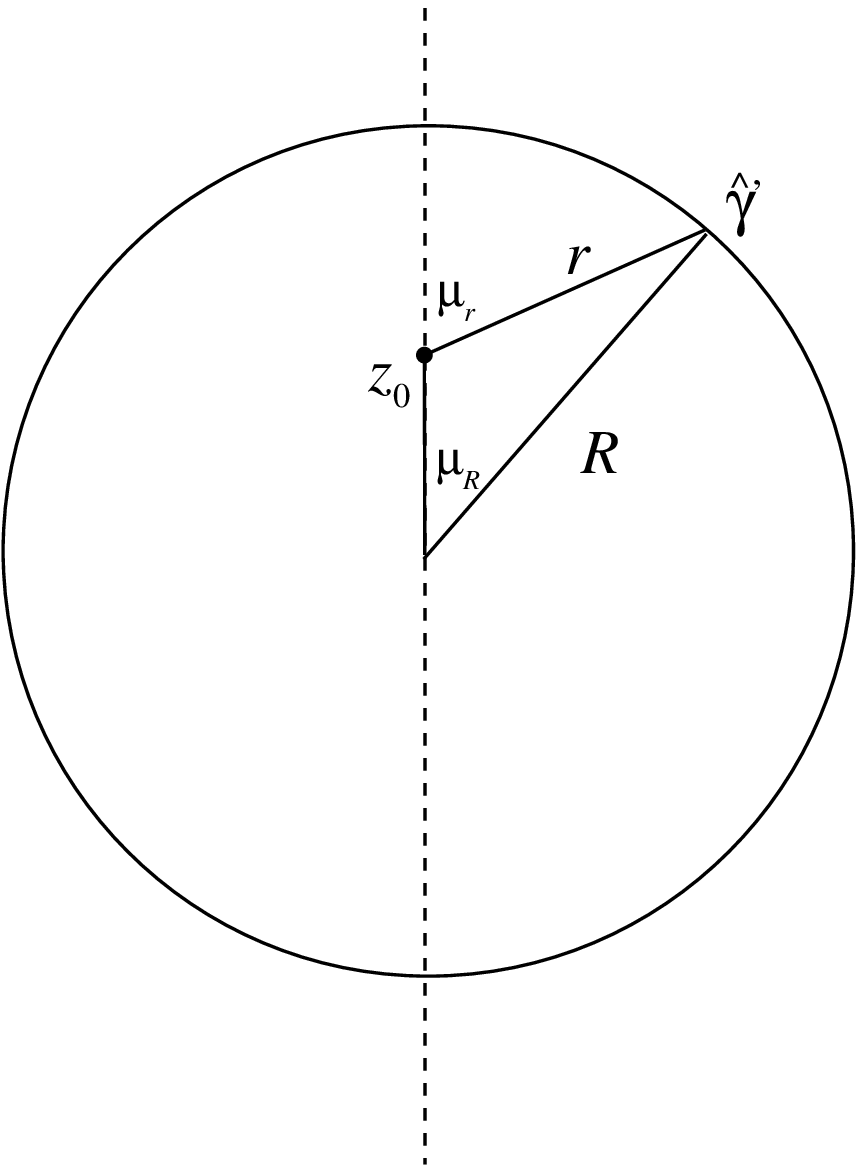}
\caption{Illustration of the geometry for a sphere with radius $R$.}
\label{fig:spherical_cloud_geom}
\end{figure}

To calculate $\gamma_{\rm T}(\vgh)$ and $\gamma_{\rm E}(\vgh)$ for spherically symmetric systems, we need the moments $y_\ell(\vek{r})$ for $\ell=0, 2$. With Eq.~\eqref{eq:Leg_y}, this boils down to computing the Legendre coefficient $y^\ast_\ell(r)\equiv  y_\ell(b=0, z)$ for the simplest case through the cluster center. Explicitly, the corresponding integral reads
\beal
\label{eq:derivation_I}
y^\ast_\ell(r)&=\frac{2\ell+1}{2}\int_{-1}^1 \int_0^\infty \frac{\sigT}{\me c^2} P_{\rm e}\left(\sqrt{r^2+s^2+2 \mu_{\rm r} s r}\right) P_\ell(\mu_{\rm r}) \id s \id \mu_{\rm r},
\end{align}
where $s$ parametrizes the integration along the direction $\vghp$ with $\mu_{\rm r}=\vgh\cdot\vghp$ (see Fig.~\ref{fig:spherical_cloud_geom} for illustration) and $P_{\rm e}(r)=k\Te(r)\Ne(r)$ is the electron pressure profile. Defining $r^\ast(s, \mu_{\rm r})=\sqrt{r^2+s^2+2 \mu_{\rm r} s r}$, Eq.~\eqref{eq:derivation_I} can be further expressed by
\beal
y^\ast_\ell(r)&=\frac{2\ell+1}{2}  \int_{0}^1 
\! P_\ell(\mu_{\rm r})
\,\frac{\sigT}{\me c^2}  
\!\int_0^\infty 
\!\Bigg\{ P_{\rm e}\left(r^\ast(s, \mu_{\rm r})\right) +P_{\rm e}\left(r^\ast(s, -\mu_{\rm r})\right)\Bigg\}  \id s \id \mu_{\rm r} 
\nonumber\\
&=\frac{2\ell+1}{2}  \int_{0}^1 P_\ell(\mu_{\rm r}) \left(\frac{\sigT}{\me c^2} \int_{-\infty}^\infty
P_{\rm e}\left(r^\ast(s, \mu_{\rm r}) \right)  \id s\right) \id \mu_{\rm r},
\end{align}
where we used that $\ell$ is even. Geometrically, the inner integral in parenthesis is simply the observed $y$ parameter, $y_{\rm obs}(b)$, as a function the effective impact parameter $b=r (1-\mu_{\rm r}^2)^{1/2}$ from the cluster center. Therefore, for even $\ell$ all $y^\ast_\ell(r)$ can be directly computed using $y_{\rm obs}(b)$.

\section{$\left< \tau^\ast_{\ell}\right>$ for simple cases}
\label{app:tau_l_simple_cases}

\subsection{Evaluation of $\left< \tau^\ast_{\ell}\right>$ for isothermal, constant density sphere}
\label{app:tau_l_const_sphere}
For illustration, we compute $\left< \tau^\ast_{\ell}\right>=\left< \tau_{\ell}\right>(b=0)$ for an isothermal, constant density sphere of electrons and line of sight directly through the center of the sphere (impact parameter $b=0$). In this case, the optical depth pattern is azimuthally symmetric around the $z$-axis which we align with $\vgh$. The optical depth in the direction $\vghp$ at location $z_0$ ($z_0=0$ at the center of the sphere) is then given by $\tau(z_0, \vghp)=\Ne \sigT r(z_0, \vghp)$, where the length through the scattering medium from the boundary of the sphere to $z_0$ is determined by $r^2=R^2+z_0^2-2 R z_0 \mu_R(\vghp)$. Here, $R$ is the radius of the sphere, and $\mu_R$ is the direction cosine towards the point on the surface of the sphere that is hit by the chosen line of sight (see Fig.~\ref{fig:spherical_cloud_geom}).
We need the Legendre coefficients, ${r}^\ast_\ell(z_0)=\frac{2\ell+1}{4\pi}\int  r(z_0, \vghp) P_\ell(\mu_r)\id \mu_r \id \varphi_{\rm r}$, where $\mu_r$ is the direction cosine for $\vghp$ relative to a coordinate system which is centered at $z_0$. Therefore, we have to express $\mu_R$ in terms of $\mu_r$. Using $R\mu_R=r\mu_r+z_0$, this eventually yields $r(\mu_r)=\left[R^2-z_0^2(1-\mu_r^2)\right]^{1/2}-z_0\mu_r$. For $\ell=0$ and $\ell=2$, we find:
\beal
{r}^\ast_0(z_0) &= \frac{R}{2}\left[1+\left(\frac{1}{\zeta_0}-\zeta_0\right){\rm arcsinh}\left(\frac{\zeta_0}{\sqrt{1-\zeta_0^2}}\right)\right]
\nonumber\\[-0.5mm]
{r}^\ast_2(z_0) &= - \frac{3R}{16}\left[1-\frac{3}{\zeta_0^2}+\left(\frac{3}{\zeta_0^3}-\frac{2}{\zeta_0}-\zeta_0\right){\rm arcsinh}\left(\frac{\zeta_0}{\sqrt{1-\zeta_0^2}}\right)\right],
\nonumber
\end{align}
with $\zeta_0=z_0/R$. To obtain $\left<\tau^\ast_{\ell}\right>$, we have to calculate $\left<{r}^\ast_{\ell}\right>=\int_{-R}^R {r}_{\ell}(z_0)\id z_0=R^2 \int_{-1}^1 {r}^\ast_{\ell}(\zeta_0)/R\id \zeta_0$. The integrals can be carried out numerically. Defining $\tau_{\rm c}=2R\Ne \sigT$ we obtain:
\beal
\label{eq:r_l_transforms}
\left<\tau^\ast_{\ell}\right>  &= (2\ell+1)\mathcal{I}_\ell\, \frac{\tau_{\rm c}^2}{2} 
\end{align}
with $\mathcal{I}_0\approx 0.867$, $\mathcal{I}_2=\pot{2.92}{-2}$, $\mathcal{I}_4=-\pot{1.80}{-3}$ and $\mathcal{I}_6=\pot{3.43}{-4}$.
The non-vanishing average optical depth multipoles drop rapidly with $\ell$. This behavior depends critically on the geometry and density profile of the scattering cloud, and for an isothermal $\beta$-profile the situations is already different (see Section~\ref{sec:iso_beta_tau}). For general impact parameter see Fig.~\ref{fig:off_center}.

\subsection{Evaluation of $\left< \tau^\ast_{\ell}\right>$ for isothermal $\beta$-profile}
\label{sec:iso_beta_tau}
As a second simple example, we consider an isothermal $\beta$-profile \citep{Cavaliere1978}:
\beal
\label{sec:Ne_prof_iso}
\Ne(r)&=N_{\rm e,0}[1+(r/r_{\rm c})^2]^{-3\beta/2}
\quad{\rm and} 
\quad
\Te\equiv {\rm const}.
\end{align}
Here, $N_{\rm e,0}\simeq 10^{-3} {\rm cm^{-3}}$ is the typical central number density of free electrons, $r_{\rm c}\simeq 100\,{\rm kpc}$ is the typical core radius of clusters, and $\beta \simeq 2/3$ \citep[e.g., see][]{Reese2002}.

The line of sight optical depth at impact parameter $b$ is 
\beal
\tau(b)&=\frac{N_{\rm e,0} \sigT r_{\rm c}}{[1+(b/r_{\rm c})^2]^{3\beta/2-1/2}}\,
\frac{\sqrt{\pi} \,\Gamma\left(3\beta/2-1/2\right)}{\Gamma(3\beta/2)},
\end{align}
which implies $\tau(b)\approx 3.142\, N_{\rm e,0} \sigT r_{\rm c}[1+(b/r_{\rm c})^2]^{-1/2}$ for $\beta=2/3$.
With the same arguments and definitions as above we find
\beal
{\tau}^\ast_{\ell}(z_0)&=
\frac{2\ell+1}{2}\int_{-1}^1 \int_0^\infty \Ne\left(\sqrt{z^2_0+s^2+2 \mu_{\rm r} s z_0}\right)\sigT P_\ell(\mu_{\rm r}) \id s \id \mu_{\rm r}.
\end{align}
For general value of $\beta$, the required integrals are best carried out numerically. For $\beta=2/3$, $\ell=0$ and $\ell=2$, we have
\beal
{\tau}^\ast_{0}(z_0)/N_{\rm e,0} \sigT r_{\rm c} &= \frac{\pi}{2} \frac{\arctan\zeta_0}{\zeta_0}
\nonumber\\[-0.5mm]
{\tau}^\ast_{2}(z_0)/N_{\rm e,0} \sigT r_{\rm c} &= \frac{3\pi}{8} \frac{(3+\zeta_0^2)\arctan\zeta_0-3\zeta_0 }{\zeta_0^3}
\nonumber
\end{align}
with $\zeta_0=z_0/r_{\rm c}$. Averaging over $\zeta_0$ (impact parameter $b=0$), we find $\left<\tau^\ast_{0}\right>=\ln 2 \times (\tau^2/2)\approx 0.693 (\tau^2/2)$, $\left<\tau^\ast_{2}\right>= 5(3-\ln 16)/8\times (\tau^2/2)\approx 0.142 (\tau^2/2)$, $\left<\tau^\ast_{4}\right>= 9(48 \ln 4-65)/256\times (\tau^2/2)\approx 0.0542 (\tau^2/2)$, where $\tau=\tau(b=0)$ is the optical depth through the center of the cluster.
Figure~\ref{fig:off_center_iso_beta} shows the non-vanishing coefficients for $\ell\leq 6$, $\beta=2/3$ and general impact parameter. 

\end{appendix}

\end{document}